# Interplay of Ni Order and Local Moments in Intercalated $Ni_xNbS_2$

Gaurav Pransu[1], Naveen Singh Dhami[1,2,3], Wojciech Sas[1,4], Yuki Utsumi Boucher[1], Bruno Gudac[1], Ana Smontara[1], Davor Tolj[5,6], Victor Balédent[2,7], Mario Novak[5], Neven Barišić[5,8], Ivo Batistić[5], Mirta Herak[1], Petar Popčević[1,*]

[1]Institute of Physics, Bijenička cesta 46, HR-10000 Zagreb, Croatia
[2]Université Paris-Saclay, CNRS, Laboratoire de Physique des Solides, Orsay, France
[3]Synchrotron SOLEIL, L'Orme des Merisiers, Saint Aubin BP 48, 91192 Gif-sur-Yvette, France
[4]Institute of Nuclear Physics Polish Academy of Sciences, Radzikowskiego 152, 31-342 Kraków, Poland
[5]Department of Physics, Faculty of Science, University of Zagreb, 10000 Zagreb, Croatia
[6]Institute for Quantum Matter, William H. Miller III Department of Physics and Astronomy, Johns Hopkins University, 3400 N Charles St, Baltimore, MD 21218, USA
[7]Institut Universitaire de France (IUF), Paris, France
[8]Institute of Solid State Physics, TU Wien, 1040 Vienna, Austria

[*]**e-mail:** ppopcevic@ifs.hr

## ABSTRACT

Magnetic ions intercalated into the van der Waals gaps of layered transition metal dichalcogenides provide a system where local moments interact strongly with a metallic host. How the order of the intercalant sublattice affects their magnetism is important but rarely addressed, although disorder is hard to avoid, in particular in thin films and heterostructures. Here this question is examined in single crystals of $Ni_xNbS_2$ with $0.10 \leq x \leq 0.57$, whose composition and Ni order were determined by energy-dispersive x-ray spectroscopy and synchrotron x-ray diffraction. Anisotropic magnetic susceptibility measurements reveal three regimes. Long-range antiferromagnetic order appears only for $0.25 \leq x \leq 0.33$, where the Ni ions form the 2×2 and $\sqrt{3}\times\sqrt{3}$ R30° superstructures, and $T_N$ reaches 96 K at $x = 1/3$. In this range the effective moment, ≈2.5 $\mu_B$ per Ni, agrees with DFT+$U$ calculations, in which the Ni spin is partly delocalized onto the surrounding S ligands. For $x \leq 0.18$ the dilute, randomly distributed Ni ions remain paramagnetic down to 2 K with a reduced moment. For $x \geq 0.37$, where Ni–Ni contacts are unavoidable, long-range order gives way to a short-range ordered state below 10–15 K with a further reduced moment. As calculations for ordered Ni arrangements find the Ni state nearly independent of $x$, even at $x = 1/2$, the reduction of the moment is attributed to a disorder-enabled reduction of the charge transfer from Ni to the $NbS_2$ layers. The magnetism of $Ni_xNbS_2$ is thus governed by the (dis)order of the Ni sublattice rather than by the Ni concentration alone.

## I. INTRODUCTION

Over the past decade, research on layered materials has increasingly turned to atomically thin films, obtained by exfoliation or grown directly, and to van der Waals heterostructures assembled from them [1,2]. In such systems, disorder is not a small perturbation but often determines the electronic and magnetic properties, and its control has become a central issue [3,4]. The role of (dis)order is a recurring theme in correlated materials in general, from dopant disorder and electronic inhomogeneity in the cuprates [5–7] to cation disorder in murunskite [8,9]. In the nickelates, where the occupation and symmetry of the Ni $e_g$ orbitals are central [10,11], oxygen vacancies on the inner apical sites change the ligand-hole content of the Ni–O states [12].
Intercalated TMDs are a particularly clean case for studying the role of (dis)order. The host layers are well ordered, and the choice of the intercalant, through the orbital character of its 3*d* states, together with its arrangement in the van der Waals gaps, determines both the coupling of the local moments to the conducting layers and the interactions between the intercalated ions. This gives access to a variety of complex magnetic orders and to the interplay between localized and itinerant electrons. Particularly interesting are chiral helimagnetism [13], altermagnetism [14] and antiferromagnetism with a large anomalous Hall effect [15,16], which make these materials attractive for spintronics [17].
Among the $NbS_2$ polytypes, 2H is the most widely studied. Pristine 2H-$NbS_2$ is metallic and becomes superconducting below ~6 K [18,19]. The material's properties, such as optical characteristics, band structure, lattice parameters, electrical transport, and magnetic properties, can be tuned through intercalation by other transition metals [20–22].
In the $TM_{1/3}NbS_2$ family (TM = first row transition metal), the intercalated ions occupy octahedral sites in the van der Waals gaps, form a $\sqrt{3}\times\sqrt{3}$ R30° superstructure and donate electrons to the $NbS_2$ conduction band [20,21,23]. Depending on the intercalant, a variety of complex magnetic ground states have been found [13,15,24–32] as well as evidence of strong correlation effects [33–36]. Even small deviations from this ideal composition can change the magnetic ground state [27,37,38]. However, only a few studies have tracked the effect of the intercalation degree systematically over a wide range of compositions, for $Co_xNbS_2$ [39], $Fe_xNbS_2$ [40,41] and $Fe_xNbSe_2$ [38], and recently for polycrystalline $Ni_xNbS_2$ with $x \leq 0.4$ [42], in some cases without determining the composition of the investigated crystals. To our knowledge, how the electronic states of the intercalant change when its order is lost has been studied spectroscopically only for $Co_{0.22}TaS_2$, $Fe_xNbS_2$ and $Ni_{0.19}NbSe_2$ [36,41,43].

In this study, we systematically investigate a series of Ni-intercalated 2H-$NbS_2$ ($Ni_xNbS_2$) single crystals with $0.10 \leq x \leq 0.57$, whose composition was determined on the measured crystals and which span both the ordered and the disordered regimes of the Ni sublattice, and compare them with the parent compound 2H-$NbS_2$. Our aim is to understand how the magnetic ground state and the Ni moment develop with the intercalation level, and how they are related to the arrangement of the Ni ions, whose ordering is reported separately [44]. Trying to correlate the magnetic properties with the electronic structure, we combine the susceptibility measurements with DFT+$U$ calculations. A central question is whether the effective magnetic moment of Ni is set by the Ni concentration itself or by the order of the Ni sublattice.

## II. EXPERIMENTAL AND COMPUTATIONAL METHODS

Single crystals of $Ni_xNbS_2$ with nominal compositions $0.01 \leq x \leq 0.67$ (Table 1) were grown by chemical vapor transport. Nickel powder (Alfa Aesar, 99.9 %, particle size 3–7 μm), niobium powder (Alfa Aesar, 99.8 %, particle size < 250 μm) and sulfur granules (Onyxmet, 99.95 %) were ground and sealed in evacuated quartz tubes with iodine as the transport agent. A temperature gradient of 50 K was applied along the tube, with the hot end at 950 °C. One batch was grown per composition, except for $x = 0.25$ and 0.33, for which several batches were grown and those with the highest $T_N$ are reported. The crystals are shiny black hexagonal platelets with lateral dimensions up to 1 cm and thicknesses up to ≈0.5 mm, and cleave easily along the *ab* plane.

Composition and morphology were examined with a Zeiss GeminiSEM 300 scanning electron microscope equipped with an Oxford Instruments energy-dispersive x-ray spectroscopy (EDX) detector. EDX maps (typically $100 \times 100$ μm$^2$, 20 kV) were recorded at several positions on several crystals of each batch and quantified with the AZtec software.

Magnetization was measured with a Quantum Design MPMS3 between 2 and 300 K, with the field applied parallel to the *ab* plane or to the *c* axis, using the vibrating-sample magnetometer (VSM) mode. Crystals of typically 0.4–3 mg were fixed on quartz paddles with GE-7031 varnish.

Synchrotron powder x-ray diffraction was measured on ground crystals from the same batches on the CRISTAL beamline of Synchrotron SOLEIL (Saint-Aubin, France), using a two-circle diffractometer with Mythen2 silicon-strip detectors ($\lambda = 0.6779$ Å, calibrated with $LaB_6$, $Q = 0.1$–8.4 Å$^{-1}$). Lattice parameters were obtained with Profex [45].

Density functional theory (DFT) calculations were performed with Quantum ESPRESSO [46,47] using the PBEsol exchange–correlation functional [48], ultrasoft pseudopotentials from pslibrary 1.0.0 [49], Marzari–Vanderbilt smearing of 0.01 Ry and the simplified rotationally invariant DFT+$U$ scheme [50] with $U = 7.0$ eV for the Ni 3$d$ states, close to the value $U = 7.7$ eV obtained for $x = 1/3$ from linear-response theory based on density-functional perturbation theory [51–53]. Plane-wave cutoffs were 100–110 Ry for the wave functions and 900 Ry for the charge density, with $k$-point meshes of 5×5×5 (3×3 supercell), 11×11×6 ($x = 1/4$), 12×12×6 ($x = 1/3$) and 20×12×6 ($x = 1/2$). Lattice parameters and atomic positions were fully relaxed. Atomic charges, orbital occupations, spin polarizations ($n\uparrow - n\downarrow$, i.e., the spin moment in $\mu_B$) and partial densities of states were obtained from a Löwdin projection of the Kohn–Sham states onto atomic orbitals [54,55].

## III. RESULTS

### A. Composition

In Fig. 1, the nominal Ni concentration used for each synthesis batch is compared with the composition determined by EDX analysis. The corresponding numerical values are summarized in Table 1. The lowest realized Ni concentration was approximately 10%, despite a nominal starting composition of 1%. With increasing nominal Ni content, the deviation from the nominal-to-realized composition decreased, and in the range $0.25 \lesssim x \lesssim 0.33$, the nominal and measured compositions were in good agreement. For nominal compositions x > 0.33, the measured Ni content fell below the nominal starting value. The full symbols in Fig. 1 (a) represent a series of syntheses performed under nearly identical growth conditions, including temperature, temperature gradient, starting material quantities, and size of the quartz ampoule, optimized for the growth of $Ni_{0.33}NbS_2$. The open symbols correspond to subsequent syntheses carried out using different amounts of starting materials and a different furnace, likely resulting in slight variations in the temperature gradient and absolute growth temperature. As a result, these two samples, with $x$ = 0.30 and 0.37, deviate more from their nominal compositions than the samples of the main series.

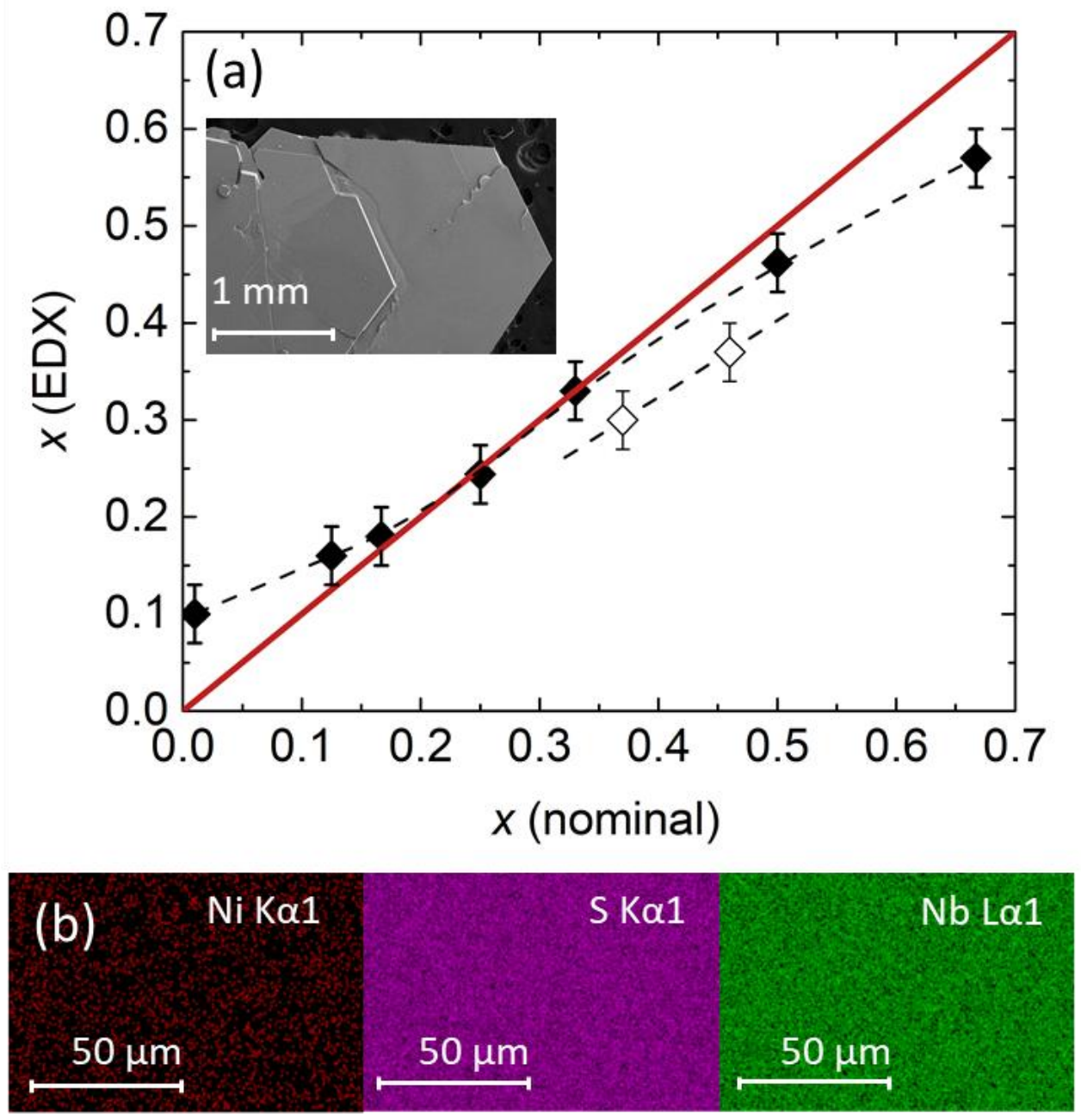

Fig. 1. (a) Comparison between the nominal Ni composition (x) used in the synthesis and the composition determined by EDX analysis. The red solid line represents the identity relation ($x_{EDX} = x_{nominal}$). Open symbols correspond to synthesis attempts performed under different conditions. Dashed lines are guides to the eye. The inset shows an SEM image of a crystal with $x_{EDX} = 0.18$. (b) EDX maps of Ni, S and Nb recorded on a part of its surface, showing no Ni-rich or Ni-poor regions on the scale of the map.

EDX compositions were obtained from maps at a minimum of two spots on each crystal, as illustrated in Fig. 1(b) for $x = 0.18$. The compositions measured at different positions agreed within the estimated uncertainty of 2 at% across the whole series. The sulfur content was consistently underestimated, primarily due to sulfur volatilization under the electron beam and the proximity of the main sulfur peak to the niobium peak.

### B. Crystal structure

The crystal structure of the series is reported in Ref. [44]. Here we summarize only the results needed for the discussion of the magnetic properties. The Ni ions occupy octahedral sites in the van der Waals gaps of 2H-$NbS_2$ ($P6_3/mmc$) [Fig. 2(a)]. Ordered superstructures are expected at $x$ = 1/4 (2×2, $P6_3/mmc$) and $x$ = 1/3 (√3×√3 R30°, $P6_322$) [23,38,40].

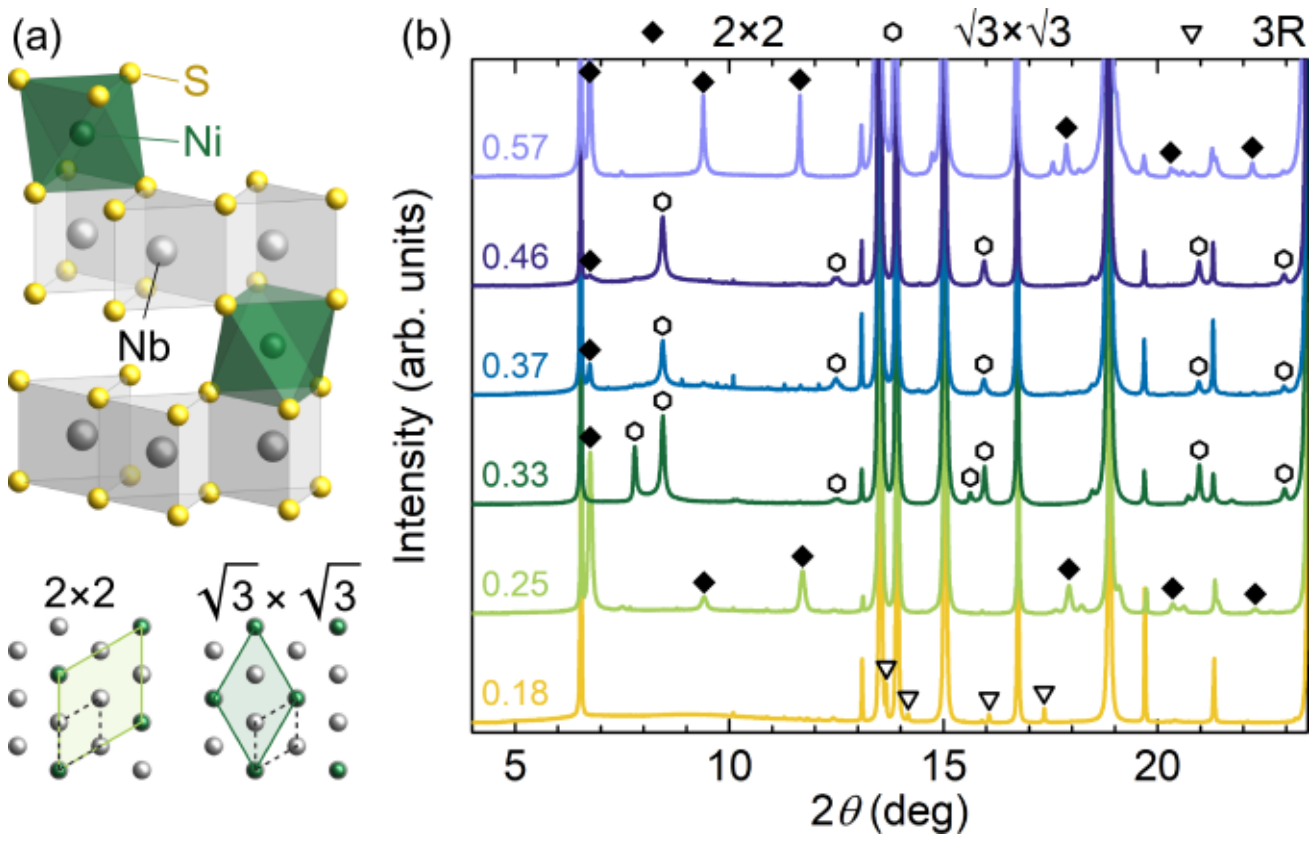

Fig. 2. (a) Schematic crystal structure of $Ni_xNbS_2$. Top: oblique side view of two $NbS_2$ layers of edge-sharing $NbS_6$ trigonal prisms (gray), with Ni (green) intercalated into octahedral sites of the van der Waals gaps. Each $NiS_6$ octahedron lies directly above or below a Nb ion. Bottom: top view of a single Ni layer for the 2×2 ($x$ = 1/4) and √3×√3 R30° ($x$ = 1/3) Ni orderings, with the Ni ions above the Nb ions of the layer below (gray). S atoms are omitted for clarity. Solid lines outline the supercells, and the dashed line marks the 2H-$NbS_2$ host unit cell. (b) Synchrotron x-ray powder diffraction patterns of $Ni_xNbS_2$ for selected $x$ ($\lambda$ = 0.6779 Å, room temperature), background-subtracted, normalized to the total integrated intensity between 4.5° and 39.5°, and vertically offset. Symbols mark the strongest superstructure reflections of the 2×2 (◆) and √3×√3 R30° (○) Ni orderings and the reflections of the 3R-$NbS_2$ secondary phase (▽).

Powder x-ray diffraction (XRD) patterns for selected values of $x$ are shown in Fig. 2. For all compositions, the diffractograms are dominated by reflections of the 2H-$NbS_2$ host. The $x$ = 0.18 crystals show no superstructure reflections, so that the Ni ions are randomly distributed. The $x$ = 0.25 sample shows only the reflections of the 2×2 superstructure, and the $x$ = 0.33 sample predominantly those of the √3×√3 R30° superstructure. For $x$ = 0.37 and 0.46, weaker 2×2 reflections appear in addition to the √3×√3 R30° reflections, so that both motifs coexist. Such motifs may develop locally even in off-stoichiometric intercalated dichalcogenides [56]. In these samples the reflections of the √3×√3 R30° order with even $l$ are absent, whereas they are present at $x$ = 0.33 [44]. Their absence is consistent with excess Ni partially occupying the site left empty by the √3×√3 R30° order. Despite the superstructure reflections, the Ni arrangement in these samples is therefore dominated by disorder. At $x$ = 0.57 the same 2×2 reflections as at $x$ = 0.25 are observed, but they are about twice as narrow, and the host reflections reveal an orthorhombic distortion. This structure can be viewed as the Ni sublattice of $x$ = 1/4 together with a second copy shifted by $a_0$, i.e., by half a lattice vector of the 2×2 cell. The Ni ions then form parallel rows of edge-sharing $NiS_6$ octahedra, separated by empty rows. This stripe order selects one in-plane direction and distorts the lattice to orthorhombic. About 0.07 Ni per formula unit remains disordered [44]. Traces of elemental sulfur (α-$S_8$, $\lesssim$ 0.1 wt%), most likely unreacted sulfur left on the crystal surfaces after the growth, are found in most samples, and a minor 3R-$NbS_2$ fraction (< 1 vol%) at $x = 0.18$. Such 3R-like stacking faults are common in 2H-$NbS_2$ itself [57], and a mixed 2H/3R stacking was reported for $x \lesssim 0.15$ [42]. Whether the 3R fraction forms separate crystallites or stacking faults in the weakly intercalated crystals, possibly introduced by grinding, cannot be decided from the powder data. No other secondary phases are identified [44]. Both minority phases are nonmagnetic and, at these fractions, do not contribute measurably to the susceptibility.

## C. Magnetic susceptibility

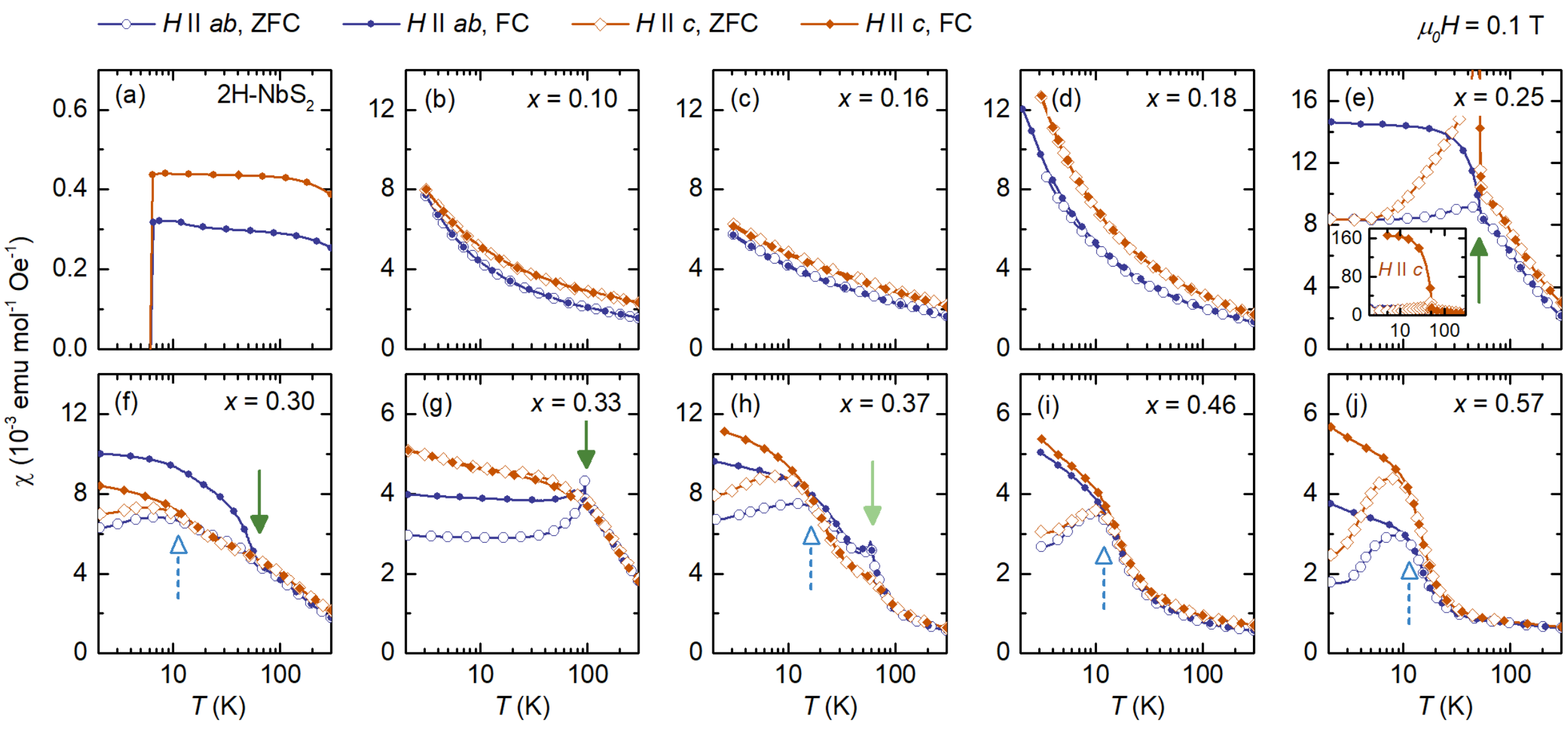

Fig. 3. Temperature dependence of the magnetic susceptibility of (a) 2H-$NbS_2$ and (b)–(j) $Ni_xNbS_2$ measured in $\mu_0 H = 0.1$ T with the field applied parallel to the *ab* plane (dark blue circles) and to the *c* axis (dark orange diamonds). Open symbols denote zero-field-cooled (ZFC) and full symbols field-cooled (FC) data. Thin lines connect all measured points, while symbols mark a subset for clarity. The temperature axis is logarithmic. The susceptibility is given per mole of Ni, except for 2H-$NbS_2$ (per mole of Nb, FC data only), where the drop below 6 K is due to the superconducting transition. Solid green arrows mark the Néel temperatures, and dashed blue arrows mark the low-temperature transitions indicated by FC-ZFC splitting (see Table 1 and text). The inset in (e) shows the full $H \parallel c$ data of x = 0.25, which exceed the main scale below $T_N$.

The temperature dependence of the magnetic susceptibility of $Ni_xNbS_2$ and of the parent compound 2H-$NbS_2$, measured in $\mu_0 H = 0.1$ T for both field orientations after zero-field cooling (ZFC) and field cooling (FC), is shown in Fig. 3. Throughout, $x$ denotes the Ni content determined by EDX (Table 1). The parent compound 2H-$NbS_2$ [Fig. 3(a)] shows a small and weakly temperature-dependent susceptibility, $\approx 2.5 \times 10^{-4}$ emu mol$^{-1}$ Oe$^{-1}$ for $H \parallel ab$ and $\approx 4 \times 10^{-4}$ emu

mol$^{-1}$ Oe$^{-1}$ for $H \parallel c$, typical of a nonmagnetic metal, and the superconducting transition at ≈6 K. Intercalation of Ni adds a paramagnetic contribution which, at low temperatures, exceeds the host susceptibility by more than an order of magnitude.

According to the low-temperature behavior, the samples fall into three groups. For $x \leq 0.18$ [Fig. 3(b)–(d)], the susceptibility increases monotonically on cooling down to the lowest temperature, without any sign of magnetic order and without FC–ZFC splitting. It is slightly larger for $H \parallel c$ than for $H \parallel ab$. For $0.25 \leq x \leq 0.33$ [Fig. 3(e)–(g)], a clear anomaly marks the Néel temperature $T_N$ (solid arrows). Below $T_N$, the susceptibility does not drop to zero for either field direction, and the FC and ZFC curves split. The splitting is particularly large for $x = 0.25$ with $H \parallel c$, where the FC susceptibility exceeds the ZFC value by more than an order of magnitude [inset of Fig. 3(e)]. For $x = 0.30$ and 0.37 [Fig. 3(f),(h)], the transition at $T_N$ is followed by a second, broader anomaly at lower temperature, below which FC and ZFC curves separate (dashed arrow). For $x = 0.46$ and 0.57 [Fig. 3(i),(j)], only this broad low-temperature anomaly remains, with FC–ZFC splitting below 10–13 K and a broad maximum of the ZFC susceptibility a few kelvin lower.

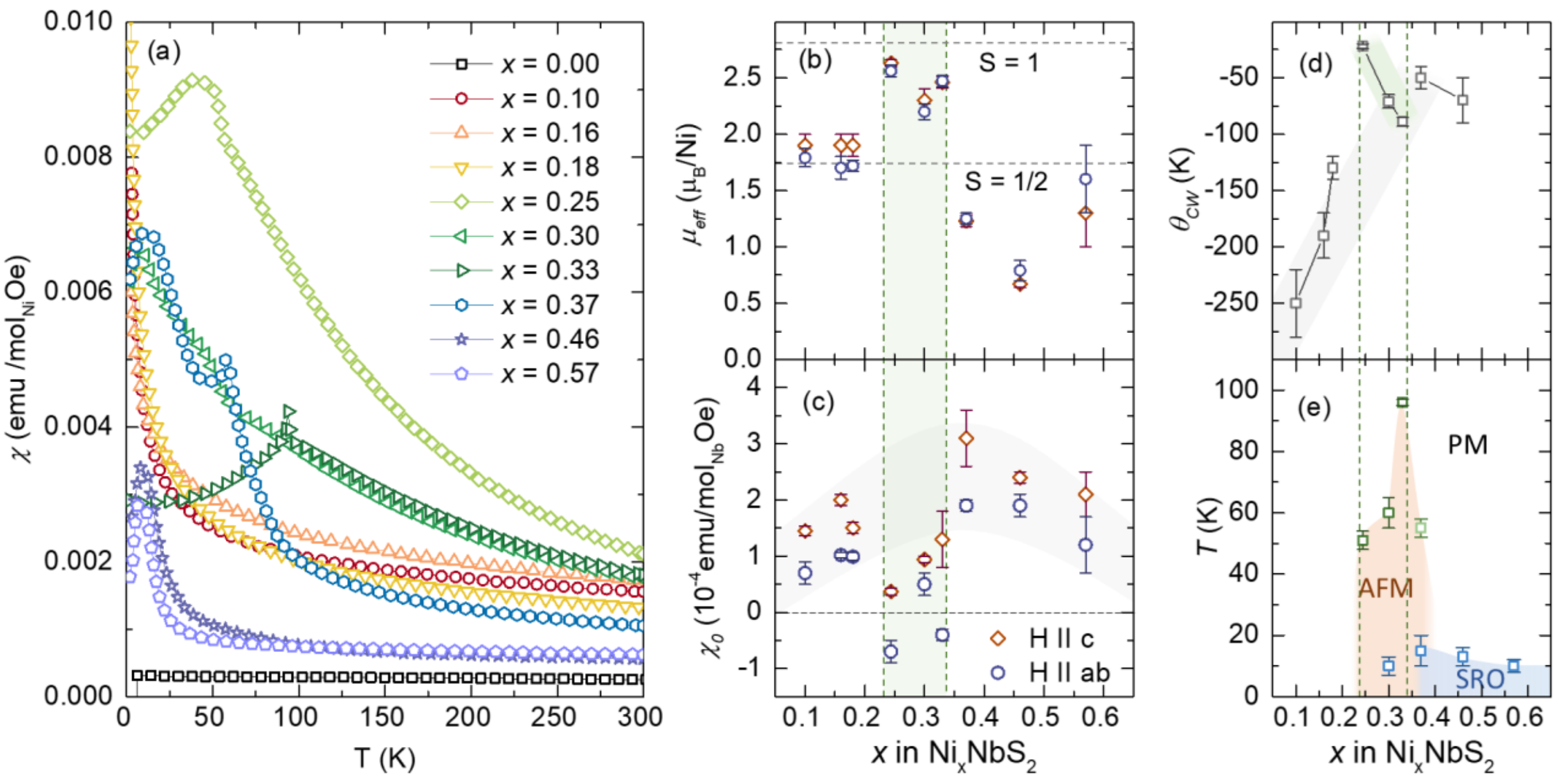

Fig. 4. (a) Magnetic susceptibility of $Ni_xNbS_2$ measured under an applied field of $\mu_0 H = 0.1$ T in ZFC conditions with the field applied parallel to the *ab* plane. The data are normalized per mole of Ni, except for x=0, where they are expressed per mole of Nb. (b) Effective magnetic moment per Ni ion as a function of composition, extracted from Curie-Weiss fits using Eq. (1). The reported values are the means obtained from measurements on multiple crystals from the same synthesis batch. Error bars indicate the standard error of the mean over the crystals of each batch. Horizontal gray dashed lines mark the expected effective moments for spin-1 and spin-1/2 states. (c) Temperature-independent magnetic susceptibility contribution obtained from Curie-Weiss fits expressed per $NbS_2$ formula unit. (d) Curie-Weiss temperature as a function of Ni composition. The value for x = 0.57, −400(100) K, lies outside the plotted range. (e) Néel temperature (green) and temperature of the low-temperature anomaly (blue) as a function

of Ni composition. Orange and blue shaded regions indicate antiferromagnetic (AFM) and short-range ordered (SRO) ground states, and PM the paramagnetic state. The gray shaded bands highlight the general trends, while the green shaded region bounded by vertical green dashed lines marks the range of the ordered superstructures, $0.25 \leq x \leq 0.33$, in which the extracted parameters deviate from these trends. The shaded regions are guides to the eye.

Fig. 4(a) shows that curves for $0.25 \leq x \leq 0.33$ stand out, featuring the highest susceptibility and the steepest temperature dependence. To analyze the paramagnetic regime quantitatively, we note that for all intercalated samples the susceptibility follows the Curie-Weiss law at sufficiently high temperatures:

$$\chi = \frac{C_{mol}}{T-\theta_{CW}} + \chi_0 \qquad (1)$$

where $C_{mol}$ is the molar Curie constant, $\theta_{CW}$ the Curie–Weiss temperature and $\chi_0$ is a temperature-independent term. It comprises the core (Larmor) diamagnetism and the response of the conduction electrons, i.e., the Pauli paramagnetism, the Landau diamagnetism and possible orbital (Van Vleck) contributions. Fits were performed above 200 K. Several crystals from each batch were measured, and the error bars in Fig. 4 represent the standard error of the mean over all crystals measured in the batch. The fit parameters are listed in Table 1 and shown in Fig. 4(b)–(d).

The effective magnetic moment per Ni is shown in Fig. 4(b). Within the green-shaded range ($0.25 \leq x \leq 0.33$) it is close to the spin-1 value, whereas outside of this range it is close to or below the spin-1/2 value. The anisotropy of magnetic moments extracted from the paramagnetic regime is within the error bars. The temperature independent contribution $\chi_0$ (Fig. 4(c)) shows a dome like dependence on $x$ (gray band), from which the data in the green-shaded range deviate, becoming even negative for H ∥ *ab*. $\chi_0$ is anisotropic, with values for H ∥ *c* consistently larger, by a similar amount in all samples, than for H ∥ *ab*. The parent compound 2H-$NbS_2$ shows the same anisotropy. No significant anisotropy is found in $\theta_{CW}$, and only the average value is shown in Fig. 4(d).

The phase diagram is shown in Fig. 4(e). Magnetic order first appears at $x = 0.25$ with Néel temperature ($T_N$) of 51 K. $T_N$ reaches its maximum of 96 K at $x = 0.33$ (compared to 84 K [28] and 90 K [20] for $Ni_{1/3}NbS_2$, and 78.5 K for polycrystalline $Ni_{0.30}NbS_2$ [42] reported earlier). At $x = 0.37$, $T_N$ decreases to 55 K and disappears after annealing (Appendix A). In the ordered state, the susceptibility does not approach zero for either field direction, which suggests that the moments lie in the *ab* plane, distributed over symmetry-equivalent domains. The ordered state is antiferromagnetic, but a weak ferromagnetic component indicated by FC–ZFC splitting is present for compositions with $x \geq 0.25$ (Fig. 3). The anomaly at about 10 K already discernible at $x = 0.30$ and consistently present for $x \geq 0.37$ is characterized by field dependence (see Appendix B).

Table 1. Composition, lattice parameters and magnetic parameters of $Ni_xNbS_2$

| EDX $x$ | Nominal $x$ | $a$ (Å)[b] | $c$ (Å)[b] | $T_N$ (K) | $T_{SRO}$ (K) | $\theta_{CW}$ (K) | $\mu_{eff}$ ($\mu_B$/Ni) | | $\chi_0$ ($10^{-4}$emu/mol Oe)[a] | |
|---|---|---|---|---|---|---|---|---|---|---|
| | | | | | | | $H \parallel ab$ | $H \parallel c$ | $H \parallel ab$ | $H \parallel c$ |
| 0.10 | 0.01 | – | – | – | – | -250(30) | 1.79(8) | 1.9(1) | 0.75(4) | 1.45(8) |
| 0.16 | 0.125 | – | – | – | – | -190(20) | 1.7(1) | 1.9(1) | 1.06(5) | 2.0(1) |
| 0.18 | 0.167 | 3.3225 | 11.8850 | – | – | -130(10) | 1.71(5) | 1.9(1) | 0.98(7) | 1.5(1) |
| 0.25 | 0.25 | 3.3226 | 11.8648 | 51(3) | – | -22(2) | 2.56(5) | 2.62(3) | -0.7(2) | 0.36(6) |
| 0.30 | 0.37 | – | – | 60(5) | 10(3) | -71(6) | 2.15(7) | 2.3(2) | 0.6(2) | 0.9(1) |
| 0.33 | 0.33 | 3.3255 | 11.8924 | 96(1) | – | -89(4) | 2.46(5) | 2.45(5) | -0.4(1) | 1.3(5) |
| 0.37 | 0.46 | 3.3276 | 11.8939 | 55(3) | 15(5) | -50(10) | 1.13(5) | 1.09(3) | 1.8(1) | 3.2(6) |
| 0.46 | 0.5 | 3.3276 | 11.8940 | – | 13(3) | -70(20) | 0.75(9) | 0.59(5) | 1.8(1) | 2.46(2) |
| 0.57 | 0.67 | 3.3344 | 11.8965 | – | 10(2) | -400(100) | 1.6(3) | 1.3(3) | 1.2(5) | 2.0(4) |

[a] $\chi_0$ values are given per mole of Nb

[b] Room-temperature lattice parameters of the parent hexagonal cell from synchrotron powder diffraction on crystals of the same batches [44]. For $x$ = 0.57 the hexagonal average of the orthorhombic cell is given.

**D. DFT results**

To relate the magnetic properties to the electronic structure, DFT+$U$ calculations were performed for $Ni_xNbS_2$ with $x$ = 1/9, 1/4, 1/3 and 1/2 (Table 2). For $x$ = 1/9, 1/4 and 1/3 the Ni ions form triangular 3×3, 2×2 and √3×√3 superstructures, respectively, so that no Ni ion has an Ni nearest neighbor. For $x$ = 1/2 we used an orthorhombic $a \times b$ cell ($b \approx \sqrt{3}a$) in which the Ni ions form parallel rows of edge-sharing $NiS_6$ octahedra along $a$, stacked on top of each other in adjacent gaps. This arrangement corresponds to the stripe order found for $x$ = 0.57 (Sec. III.B). In all calculations the Ni moments are coupled ferromagnetically within a layer and antiferromagnetically between adjacent layers.

The relaxed lattice parameters (Table 2) agree with the experimental ones (Table 1) to within 0.5 %, and for $x$ = 1/2 the calculation also reproduces the orthorhombic distortion found for $x$ = 0.57.

The electronic state of the Ni ion is remarkably insensitive to the Ni concentration (Table 2). The $NiS_6$ octahedron is slightly elongated along $c$, so that the local symmetry is trigonal antiprismatic. In the trigonal frame ($z \parallel c$) the octahedral $t_{2g}$ level splits into $a_{1g}$ ($d_{z^2}$) and $e_g^{\pi}$, while the octahedral $e_g$ pair, whose lobes point toward the S ligands and which is σ-bonded to them, becomes $e_g^{\sigma}$. The trigonal splitting is small compared with the octahedral crystal field [56,58]. In the idealized ionic limit, for $Ni^{2+}$ ($3d^8$) all majority-spin orbitals and the minority-spin $t_{2g}$ ($a_{1g}$ and $e_g^{\pi}$) orbitals are occupied, whereas the two minority-spin $e_g^{\sigma}$ orbitals are empty. This gives $S$ = 1 and a spin moment of 2 $\mu_B$. In $Ni_xNbS_2$ this configuration is strongly modified by covalency, as shown below.

The corresponding densities of states are shown in Fig. 5. For $x \leq 1/3$ the states at $E_F$ derive predominantly from the Nb $d_{z^2}$ band, whose upper edge lies 1.3–1.6 eV above $E_F$ [Fig. 5(a)]. Within this band, 0.6–1.1 eV above $E_F$, a narrow double peak of predominantly Ni minority-spin $e_g^\sigma$ character appears, while the occupied Ni 3$d$ states lie well below $E_F$, between −7 and −3 eV [Fig. 5(b)]. In a simple picture, the electron count in this energy range follows from the configuration above. In 2H-$NbS_2$ the conducting Nb $d_{z^2}$ band holds two states per Nb and is half filled. Each Ni contributes ten valence electrons ($3d^8 4s^2$) and ten 3$d$ states. Eight 3$d$ states are occupied and the two empty minority-spin $e_g^\sigma$ states lie within the Nb band, while the two 4$s$ electrons are transferred to the Nb band. Between $E_F$ and the gap above this conduction band, formed by the Nb $d_{z^2}$ and Ni $e_g^\sigma$ states, there should therefore be $1 - 2x$ unoccupied states per Nb in the Nb band and $2x$ in the Ni peaks, i.e., one unoccupied state per Nb independent of $x$. The calculations give 0.97–1.00 unoccupied states per Nb for all $x$, with ≈2 states per Ni in the double peak. With increasing $x$ the Fermi level therefore approaches the top of the Nb band, and the density of states at $E_F$ decreases from 1.8 to 1.4 and 1.0 states $eV^{-1}$ per Nb for $x$ = 1/9, 1/4 and 1/3, respectively. For $x$ = 1/2, where in this count no empty Nb states remain, the double peak broadens and its lower edge reaches $E_F$ [Fig. 5(b)], so that the Ni contribution to the density of states at $E_F$ increases from below 7 % for $x \leq 1/3$ to ≈40 %.

The two states per Ni in the double peak are, however, not pure Ni states. Through the σ bonds the $e_g^\sigma$ orbitals hybridize with S 3$p$ and, via S, with the Nb $d_{z^2}$ band. Accordingly, the minority-spin Ni 3$d$ weight integrated from $E_F$ to the top of the conduction band, most of it in the double peak, is only 0.85–1.02 per Ni [Fig. 5(c)], the remainder of the peak being of S and Nb character. Conversely, the $e_g^\sigma$ orbitals acquire weight in occupied states, in the S 3$p$ bands and in the occupied part of the Nb band. Further unoccupied minority-spin Ni 3$d$ weight lies above the gap, where the $e_g^\sigma$ orbitals mix with the higher-lying Nb 4$d$ and S states. The total number of unoccupied minority-spin Ni 3$d$ states thus reaches 1.12–1.25 per Ni (Table 2). The majority-spin 3$d$ orbitals and the minority-spin $t_{2g}$ orbitals, in contrast, remain almost filled, with only ≈0.1 holes each per Ni. The spin polarization of Ni, i.e., the difference between the numbers of unoccupied minority- and majority-spin Ni states, is therefore ≈0.1 smaller, 1.01–1.15 $\mu_B$ (Table 2), about half of the ionic value.

This reduction does not mean that the spin is lost. Because the empty minority-spin $e_g^\sigma$ states are Ni–S σ-antibonding combinations with a sizable S 3$p$ admixture, the spin of the Ni ion extends onto the surrounding S atoms. The S atoms bonded to Ni carry 0.29–0.33 $\mu_B$ per Ni, parallel to the Ni moment, whereas the S atoms not bonded to Ni (≤0.01 $\mu_B$) and the Nb atoms (≤0.1 $\mu_B$, with alternating sign) are only weakly polarized. The ligand polarization is thus not an additional moment but part of the same spin, delocalized over the $NiS_6$ cluster, which carries 1.30–1.48 $\mu_B$ in total and is the magnetic unit relevant for the paramagnetic response. For comparison with the Curie–Weiss analysis we take $S = m/2$, where $m$ is the polarization of Ni alone or of the whole $NiS_6$ cluster, and account for the orbital contribution, which arises from spin–orbit coupling within the Ni 3$d$ shell, by $g$ = 2.2, the value found by electron spin resonance for $Ni^{2+}$ in the similarly covalent $NiS_6$ octahedra of $Ni_2P_2S_6$ [59]. This gives $\mu_{eff}$ = 1.9–2.1 $\mu_B$ for Ni alone and 2.3–2.5 $\mu_B$ for the $NiS_6$ cluster (Table 2).

At $x = 1/2$, where every Ni ion has two Ni nearest neighbors, the only qualitative difference is the position of the $e_g^\sigma$ states at $E_F$. The occupation of the Ni states, however, changes little. The Ni spin polarization decreases by less than 10 % compared to $x = 1/3$ (Table 2), mainly through a small increase of the minority-spin Ni 3*d* weight in the Ni–S bonding states well below $E_F$. The total S polarization per Ni is preserved. The S atoms shared by two Ni ions are more strongly polarized, their polarization being divided between the two Ni ions, while the S atoms bonded to a single Ni ion are less polarized than for $x \leq 1/3$.

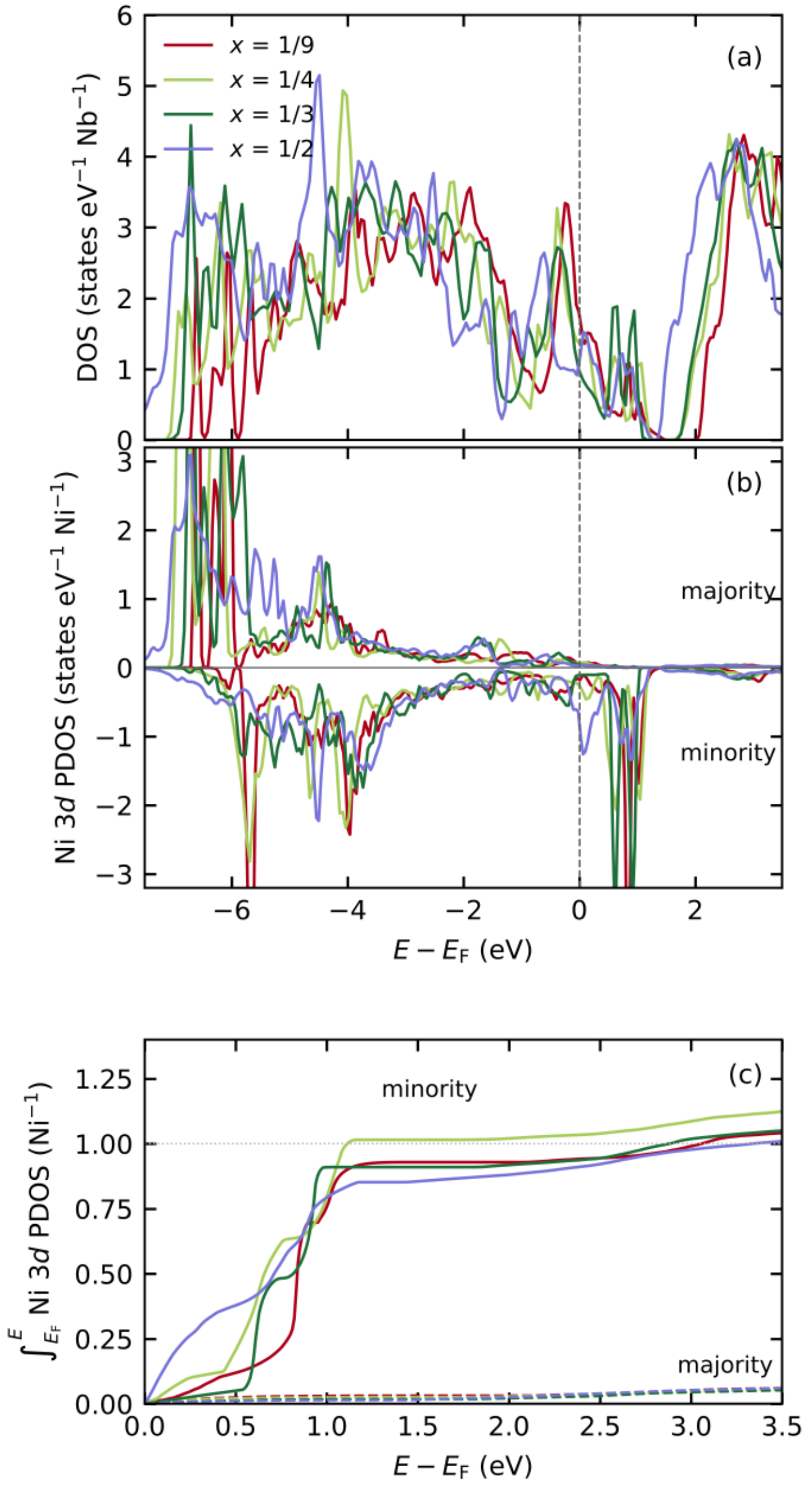

Fig. 5. (a) Total density of states per Nb and (b) spin-resolved Ni 3*d* partial density of states of one Ni ion (majority spin positive, minority spin negative) of $Ni_xNbS_2$ from DFT+*U* for $x$ = 1/9, 1/4, 1/3 and 1/2. (c)

Number of unoccupied minority-spin (solid lines) and majority-spin (dashed lines) Ni 3*d* states per Ni between $E_F$ and $E$, obtained by integrating the partial density of states in (b). The energy is measured from the Fermi level.

Table 2. Results of DFT+$U$ calculations for $Ni_xNbS_2$. Occupations and spin polarizations (in $\mu_B$) are obtained from a Löwdin projection on atomic orbitals. $a_0$ is the lattice parameter of the 1×1 hexagonal cell. $\mu_{eff} = g[S(S+1)]^{1/2}$ with $g = 2.2$ [59] and $S = m/2$, where $m$ is the polarization of Ni alone or of Ni together with that of S per Ni ($NiS_6$).

| | $x = 1/9$ | $x = 1/4$ | $x = 1/3$ | $x = 1/2$ |
|---|---|---|---|---|
| Supercell | 3×3 | 2×2 | √3×√3 | $a \times b$[a] |
| Atoms | 56 | 26 | 20 | 14 |
| $a_0$ (Å) | 3.309 | 3.313 | 3.316 | 3.322 |
| $c$ (Å) | 11.858 | 11.850 | 11.853 | 11.842 |
| Ni 3*d* occupancy | 8.69 | 8.66 | 8.70 | 8.77 |
| Empty minority-spin 3*d* | 1.21 | 1.25 | 1.20 | 1.12 |
| Ni spin polarization | 1.10 | 1.15 | 1.10 | 1.01 |
| S polarization per Ni | 0.31 | 0.33 | 0.30 | 0.29 |
| $\mu_{eff}$: Ni | 2.03 | 2.10 | 2.04 | 1.92 |
| $\mu_{eff}$: $NiS_6$ | 2.41 | 2.50 | 2.40 | 2.28 |
| $N(E_F)$ ($eV^{-1}$ $Nb^{-1}$) | 1.8 | 1.4 | 1.0 | 1.2 |

[a] Orthorhombic cell with $a \approx a_0$ along the Ni rows and $b \approx \sqrt{3}a_0$, with $a_0 = (ab/\sqrt{3})^{1/2}$.

## IV. DISCUSSION

Figure 4 illustrates the change of all parameters at the boundaries of the composition range $0.25 \leq x \leq 0.33$, in which the Ni ions form the long-range ordered 2×2 and √3×√3 R30° superstructures [44]. Outside this range the Ni subsystem is characterized by disorder, although local superstructure motifs persist (Sec. III.B). Three regimes can therefore be distinguished. In the $x \leq 0.18$ regime, the Ni ions are diluted and randomly distributed, no magnetic order is found down to 2 K, the effective moment is reduced to ≈1.8 $\mu_B$ and $\theta_{CW}$ is strongly negative. In the ordered regime, $0.25 \leq x \leq 0.33$, the effective moment is largest, 2.2–2.6 $\mu_B$, the system orders antiferromagnetically with $T_N$ up to 96 K, and $\chi_0$ is smallest. In the over-intercalated regime, $x \geq 0.37$, Ni–Ni contacts in the form of edge-sharing octahedra are unavoidable, long-range order gives way to a short-range ordered (SRO) state below 10–15 K, the effective moment drops to 0.6–1.6 $\mu_B$ and $\chi_0$ is largest. The $x = 0.30$ sample, which lies between the two superstructures, shows intermediate parameters and the low-temperature anomaly characteristic of the disordered samples. Below we first use the ordered regime, to which the DFT calculations apply most directly, as a reference and then discuss the deviations on either side of it.

In $Ni_{1/3}NbS_2$, ARPES and DFT showed that, to a first approximation, the conduction bands are those of 2H-$NbS_2$ filled by the electrons donated by Ni, while the Ni 3*d* states are pushed away from $E_F$ by strong on-site correlations [34]. The calculations of Sec. III.D confirm this picture and show that the spin is shared by the whole $NiS_6$ cluster. Its effective moment, $\mu_{eff}$ = 2.4–2.5 $\mu_B$, is in good agreement with the measured 2.45–2.6 $\mu_B$ for $x = 0.25$ and 0.33. Thus DFT+$U$ describes the ordered compounds well. The delocalization of the spin onto the S ligands also explains the much smaller ordered moment of 1.05 $\mu_B$ in $Ni_{1/3}NbS_2$ [28] observed by neutron diffraction. The Curie–Weiss constant measures the mean square of the whole fluctuating moment, $\langle\mu^2\rangle = \mu_{\text{eff}}^2\mu_B^2 = g^2S(S+1)\mu_B^2$, including its ligand part. Neutron diffraction, in contrast, measures $gS\mu_B$, and a refinement with the $Ni^{2+}$ magnetic form factor largely misses the spin density on the ligands [60]. The crystals of Ref. [28] order at a lower temperature, 84 K, which may indicate more disorder of the Ni sublattice and a further reduction of the moment. In our series, the $x = 0.30$ crystal likewise has a lower $T_N$ and a reduced moment (Table 1), consistent with all disordered samples discussed below.

The ordered regime is also where $\chi_0$ is smallest. Here $\chi_0$ is consistent with the calculated electronic structure. The core diamagnetism, $\approx -0.8\times10^{-4}$ emu mol$^{-1}$ Oe$^{-1}$ per Nb [61], and the Pauli susceptibility $\mu_B^2N(E_F)$ [62] of $0.45\times10^{-4}$ and $0.32\times10^{-4}$ emu mol$^{-1}$ Oe$^{-1}$ for $x = 1/4$ and 1/3, reduced by about one third by the Landau diamagnetism, give $\approx -0.5\times10^{-4}$ emu mol$^{-1}$ Oe$^{-1}$, close to the values measured for $H \parallel ab$. The anisotropy $\chi_0(H \parallel c) > \chi_0(H \parallel ab)$ is present already in 2H-$NbS_2$, is of similar size in all samples and does not affect the $x$ dependence of $\chi_0$ discussed below.

For $x \leq 0.18$ the effective moment, $\approx 1.8\ \mu_B$, is about 30 % smaller than in the ordered regime. A reduced moment of dilute, disordered intercalants is a common observation. In $Fe_xNbS_2$, $\mu_{\text{eff}} = 4.77\ \mu_B$ (close to the spin-only value of $Fe^{2+}$) for the ordered $x = 1/4$ composition, but only 1.76 $\mu_B$ for $x = 0.15$ [40]. In $Fe_xNbSe_2$ the moment per Fe decreases below $x = 0.15$ [38], and in $Co_xTaS_2$ $\mu_{\text{eff}}$ decreases from 3.0 $\mu_B$ at $x \approx 1/3$ to 2.4 $\mu_B$ at $x = 0.22$ [63]. In $Co_{0.22}TaS_2$ the ARPES spectra are consistent with a reduced charge transfer, and the changed shape of the Fermi surface suggests a modified orbital character of the Co states near $E_F$ [36]. In $Fe_xNbS_2$, too, the charge transferred to the Nb bands varies nonmonotonically with $x$ [41]. The suppression of the Curie moment goes together with an enhancement of $\chi_0$. The dome-like dependence of $\chi_0$ on $x$ [Fig. 4(c)] has its maxima where $\mu_{\text{eff}}$ is reduced [Fig. 4(b)]. Although the weak temperature dependence of the susceptibility of the disordered samples between 200 and 300 K leaves some freedom in how it is divided between $C_{\text{mol}}$ and $\chi_0$, this correlation suggests that the weight missing from the Curie term is transferred to the temperature-independent susceptibility. A natural way for this to happen is a reduced charge transfer.

In the ordered structures the strong on-site correlations, included in the calculations through $U$, keep the minority spin $e_g^\sigma$ states above $E_F$, and their degeneracy is protected by the threefold axis through the Ni site. Disorder in the Ni arrangement can break the threefold symmetry of the local environment, splitting and broadening the $e_g^\sigma$ doublet and bringing it toward $E_F$, allowing a smaller charge transfer. The opposite, weakly correlated limit is illustrated by the calculations of Refs. [28,42] for $Ni_{1/3}NbS_2$ and $Ni_{0.25}NbS_2$ without $U$. In these calculations the minority-spin Ni states lie at and just below $E_F$. This reduces the Ni moment and increases the Ni contribution to the density of states at $E_F$. This reduced moment is consistent with the weaker magnetism of their less ordered samples. Our disordered samples move from the DFT+$U$ picture of the ordered phases toward this limit, with a reduced moment and an enhanced $\chi_0$. Our calculation for $x = 1/2$ also indicates that a state with a reduced moment is close in energy. When constrained to the hexagonal metric, i.e., without the orthorhombic distortion, it converges to a nonmagnetic state only slightly higher in energy than the orthorhombic magnetic state (not shown). Whether the reduction is static, i.e., a distribution of Ni sites with different moments, or partly dynamic cannot be decided from the susceptibility alone. In either case it appears only where the Ni order is disturbed.

For $x \geq 0.37$ the Ni ions necessarily have Ni nearest neighbors in edge-sharing octahedra. In the calculation for $x = 1/2$ (Sec. III.D) such contacts change the local moment only slightly but bring the $e_g^{\sigma}$ states to $E_F$, so that a moderate disorder of the Ni positions may suffice to fill them partially. This reduces the spin and adds Ni weight at $E_F$, in line with the largest $\chi_0$ and lowest $\mu_{eff}$ found at higher $x$. We therefore attribute the reduced Curie constant to the same disorder-driven mechanism as at low $x$. The regimes differ in the Ni–Ni contacts, around which the remaining moments are coupled at high $x$ and form the SRO state discussed below. The moment partially recovers for $x = 0.57$, whose orthorhombic arrangement of parallel Ni rows is structurally the closest to the calculated $x = 1/2$. A similar recovery of the moment at high $x$ was reported for $Fe_xNbS_2$, where $\mu_{eff}$ decreases from 4.8 $\mu_B$ at $x = 0.25$ to 2.8 $\mu_B$ at $x = 0.33$ and increases again to 4.5 $\mu_B$ at $x = 0.50$ [40]. The transition at 55 K in $x = 0.37$, which disappears upon annealing (Appendix A), presumably stems from locally $\sqrt{3}\times\sqrt{3}$-ordered regions.

Alternatively, the Ni moments could be partially screened by the conduction electrons, as in the Kondo-like behavior reported for dilute $Fe_xNbS_2$ [64] and $Fe_{0.05}NbSe_2$ [38]. If active at high temperatures, this mechanism would also transfer weight from the Curie term to $\chi_0$ [65,66]. This would, however, require a distribution of screening scales from below 10 K (to account for the Curie tail of the susceptibility) to well above room temperature and an abrupt disappearance of the screening in the ordered regime. We therefore consider this scenario less likely. In contrast to Kondo screening, which compensates the Ni spin by the conduction electrons but leaves the Ni charge and the filling of the Nb bands unchanged, a reduced charge transfer changes the charge of the Ni ion itself and thereby lowers the band filling. ARPES can distinguish the two pictures by tracking the size of the Nb-derived Fermi pockets and the depth of the conduction bands, as done for $Fe_xNbS_2$ [41], $Co_{0.22}TaS_2$ [36] and $Ni_{0.19}NbSe_2$ [43]. Published data already point in this direction. In nominal $NiNb_3S_6$ crystals, for which neither the composition nor the ordering temperature was reported, the Γ-centered hole pocket was found to be nearly as large as in 2H-$NbS_2$, and the Nb bands extend to nearly the same depth below $E_F$ [67]. In another study where crystal quality was well established, in contrast, the conduction band is shifted toward larger binding energies, consistent with the band filling obtained from ab initio calculations [34]. A band filling that differs between nominally identical $Ni_{1/3}NbS_2$ crystals is naturally explained by a charge transfer that depends on the amount and the order of Ni, but not by Kondo screening. A similar situation is found in disordered $Ni_{0.19}NbSe_2$, where the van Hove singularity shifts by only ≈50 meV upon intercalation and the analysis of the authors points to a Ni valence between 1+ and 2+, i.e., an incomplete charge transfer [43]. Systematic ARPES measurements across the composition series would allow a direct test of this picture.

The Curie–Weiss temperature [Fig. 4(d)] is negative for all samples. Within the ordered regime $|\theta_{CW}|$ increases from $x = 0.25$ to 0.33 and thus roughly follows $T_N$. For $x \leq 0.18$, its large magnitude without order down to 2 K is unlikely to reflect exchange between the dilute Ni moments and, like the reduced moment and the enhanced $\chi_0$, suggests that the Ni moments in a disordered environment are not stable local moments. The same holds for $x$ = 0.57, where $|\theta_{CW}| \approx 400$ K exceeds the temperature of the low-temperature anomaly by more than an order of magnitude. For $x$ = 0.30 and $x \geq 0.37$, the FC and ZFC curves split below 10–15 K, and the ZFC susceptibility has a broad maximum a few kelvin lower. The maximum and the splitting are suppressed in a field of 1 T (Appendix B). We attribute this state to short-range magnetic order of ferromagnetically correlated Ni clusters, formed around Ni–Ni contacts and other defects of the Ni order and coupled antiferromagnetically to each other, and refer to it as short-range ordered (SRO). Related states, a reentrant spin glass in $Fe_{0.38}NbSe_2$ [38] and coexisting antiferromagnetic and spin-glass components in $Fe_xNbS_2$ for $x \geq 1/3$ [40], appear in other intercalated dichalcogenides away from the best-ordered compositions.

Previous studies of the $x$ dependence of intercalated $NbS_2$, $NbSe_2$ and $TaS_2$ [27,37–40,63] likewise show that the magnetism is sensitive to the intercalant content and order. Compared to other studies, we combine (i) compositions determined by EDX on the measured single crystals with the Ni order determined by synchrotron x-ray diffraction [44]; (ii) a wide range, $0.10 \leq x \leq 0.57$, on both sides of the ordered phases, with the highest $T_N$ reported so far (96 K) at $x$ = 1/3 indicating a well-ordered Ni sublattice; (iii) an anisotropic Curie–Weiss analysis of several crystals per batch; and (iv) DFT calculations for ordered Ni arrangements spanning the same range, including Ni–Ni contacts for $x$ = 1/2.

## V. CONCLUSION

We have studied single crystals of $Ni_xNbS_2$ with $0.10 \leq x \leq 0.57$, whose composition was determined by EDX on the measured crystals and whose Ni order was determined by synchrotron x-ray diffraction, by anisotropic magnetic susceptibility, and by DFT+$U$ calculations. Long-range antiferromagnetic order exists only for $0.25 \leq x \leq 0.33$, where the Ni ions form the 2×2 and $\sqrt{3}\times\sqrt{3}$ R30° superstructures, with $T_N$ reaching 96 K at $x$ = 1/3. In this long-range ordered regime, the effective magnetic moment, $\mu_{eff} \approx 2.5\ \mu_B$ per Ni, agrees with DFT+$U$. In the calculations the Ni spin is partly delocalized onto the six S ligands, so that the magnetic unit is the $NiS_6$ cluster. Outside of this regime, the Ni sublattice is dominated by disorder. As a result, the effective moment is reduced and the temperature-independent susceptibility is enhanced. For $x \leq 0.18$ no magnetic order is found down to 2 K, while for $x \geq 0.37$ long-range order is replaced by a short-range ordered state below 10–15 K. Because calculations for ordered Ni arrangements give a nearly $x$-independent Ni state, even with Ni–Ni contacts at $x$ = 1/2, we attribute these changes to disorder, which weakens the effect of the strong on-site correlations and allows a reduced charge transfer from Ni to the $NbS_2$ layers. Kondo-like screening of the Ni moments is a less likely alternative. Unlike Kondo screening, which compensates the Ni spin but leaves the filling of the Nb bands unchanged, a reduced charge transfer lowers the band filling, and published ARPES data favor this scenario. The magnetism of $Ni_xNbS_2$ is thus controlled to a large extent by the order of the Ni sublattice rather than by the Ni concentration alone. This is relevant beyond bulk crystals. Thin films and heterostructures of intercalated TMDs are now prepared by intercalating thin flakes or patterned devices [68], and in such samples the intercalant sublattice is rarely well ordered. Controlling the order of the intercalant is therefore as important as controlling its amount.

Acknowledgment

This work has been supported by the Croatian Science Foundation under the project numbers IP-2020-02-9666, IP-2025-02-7384. G. P, W. S., M. H. and P. P. acknowledge support of the project *Cryogenic Centre at the Institute of Physics – KaCIF* co-financed by the Croatian Government and the European Union through the European Regional Development Fund-Competitiveness and Cohesion Operational Programme (Grant No. KK.01.1.1.02.0012) and the project *Ground states in competition – strong correlations, frustration and disorder — FrustKor* financed by the Croatian Government and the European Union through the National Recovery and Resilience Plan 2021-2026 (NRRP). We acknowledge the SOLEIL Synchrotron and the scientists of the CRISTAL beamline for providing beamtime (proposal number 20251118).

## APPENDIX A: ANNEALING OF $Ni_{0.37}NbS_2$

Figure A1 compares the susceptibility of an as-grown crystal with $x = 0.37$ with that of the same crystal after annealing for 10 min at 200 °C. The as-grown crystal shows the antiferromagnetic transition at $T_N \approx 55$ K and, at lower temperatures, FC–ZFC splitting. After annealing, the anomaly at $T_N$ is absent and only the broad low-temperature anomaly remains, with a maximum of the ZFC susceptibility at ≈10 K and FC–ZFC splitting below ≈15 K. The annealed crystal thus behaves like the crystals with $x = 0.46$ and 0.57. This supports the interpretation that the transition at $T_N$ stems from locally $\sqrt{3}\times\sqrt{3}$-ordered regions (Sec. IV), which are destroyed already by a mild annealing.

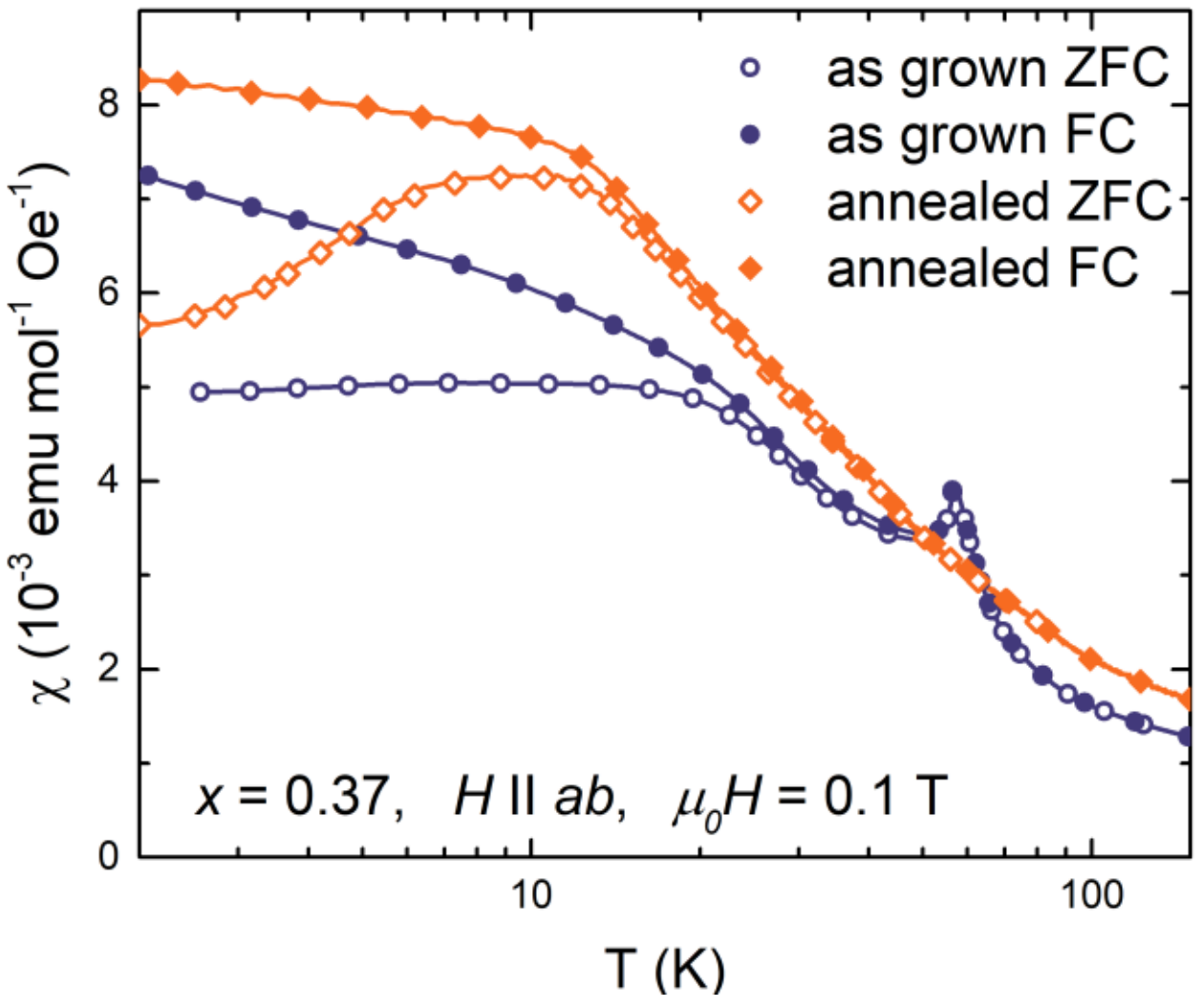

Fig. A1. Temperature dependence of the magnetic susceptibility of $Ni_{0.37}NbS_2$ per mole of Ni, measured in $\mu_0 H = 0.1$ T with $H \parallel ab$ after ZFC (open symbols) and FC (full symbols), for the as-grown crystal (dark blue circles) and after annealing for 10 min at 200 °C (orange diamonds). The temperature axis is logarithmic.

## APPENDIX B: SHORT-RANGE ORDERED STATE

Figure B1 shows the ZFC and FC susceptibility of $x = 0.46$ in 0.1 and 1 T ($H \parallel ab$). In 0.1 T the FC and ZFC curves split below ≈13 K, and the ZFC susceptibility has a broad maximum at ≈10 K. In 1 T both the maximum and the splitting are absent.

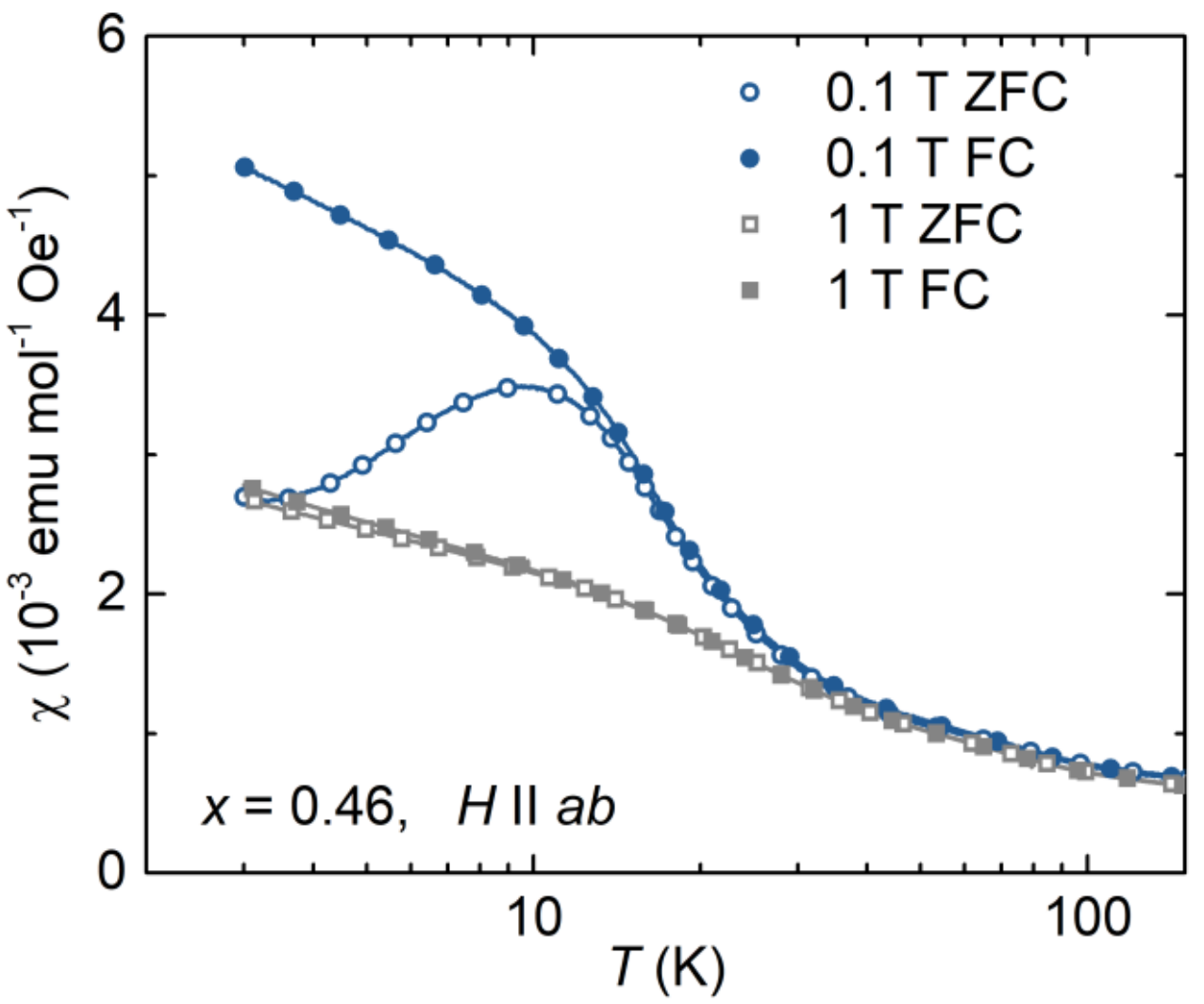

Fig. B1. Temperature dependence of the molar magnetic susceptibility ($\chi_{molar}$) of $Ni_{0.46}NbS_2$ measured after ZFC (open symbols) and FC (full symbols) in 0.1 T and 1 T parallel to the $ab$-plane (H ∥ ab). The temperature axis is logarithmic.